\documentclass{amsart}
\usepackage{amsthm,amssymb}
\usepackage{enumerate}

\newtheorem{example}{Example}
\newtheorem{theorem}{Theorem}
\newtheorem{lemma}{Lemma}
\newtheorem{answer}{Answer}
\newtheorem{algorithm}{Algorithm}
\newtheorem{definition}{Definition}

\newtheorem{exercise}{Exercise}

\usepackage{xspace}
\DeclareMathOperator{\dlog}{dlog}
\DeclareMathOperator{\poly}{poly}
\DeclareMathOperator{\id}{id}

\DeclareMathOperator{\aut}{Aut}

\newcommand{\C}{\mathbb{C}}
\newcommand{\F}{\mathbb{F}}
\newcommand{\N}{\mathbb{N}}
\renewcommand{\P}{\mathbb{P}}
\newcommand{\Q}{\mathbb{Q}}
\newcommand{\R}{\mathbb{R}}
\newcommand{\Z}{\mathbb{Z}}
\newcommand{\e}{\mathrm{e}}
\renewcommand{\i}{\mathrm{i}}
\newcommand{\HSP}{\textsc{hsp}\xspace}
\newcommand{\ket}[1]{|#1\rangle}

\begin{document}

\title[Quantum algorithms for problems in NT, Ag, and GT]{Quantum algorithms for problems in number theory, algebraic geometry, and group theory}

\author{Wim van Dam}

\address{Wim van Dam $\cdot$ Department of Computer Science, Department of Physics, University of California, 
Santa Barbara, California, 93106-5110, USA $\cdot$
{vandam@cs.ucsb.edu}}

\author{Yoshitaka Sasaki}
\address{Yoshitaka Sasaki $\cdot$
Osaka University of Health and Sport Sciences, 1-1 Asashirodai, Kumatori-cho, Sennan-gun, 
Osaka, 590-0496, Japan $\cdot$ ysasaki@ouhs.ac.jp}

\begin{abstract}
Quantum computers can execute algorithms that sometimes dramatically outperform classical computation. 
Undoubtedly the best-known example of this is Shor's discovery of an efficient quantum algorithm for
 factoring integers, whereas the same problem appears to be intractable on classical computers. Understanding what other 
computational problems can be solved significantly faster using quantum algorithms is one of the 
major challenges in the theory of quantum computation, and such algorithms motivate the 
formidable task of building a large-scale quantum computer. This article will review the 
current state of quantum algorithms, focusing on algorithms for problems with an algebraic flavor that 
achieve an apparent superpolynomial speedup over classical computation.
\end{abstract}


\maketitle
\thispagestyle{empty}

\newpage
\tableofcontents
\newpage
\section{Introduction}
Ever since Shor~\cite{sh} showed in 1994 that a quantum computer can efficiently
factor integers and calculate discrete logarithms, many more quantum algorithms have been discovered. 
In this article, we review some quantum algorithms and the problems that they try to solve such as the discrete logarithm problem, the Abelian and non-Abelian hidden subgroup problem, the point counting problem of finite field equations, 
and so on. 
A significant number of these algorithms follow the set-up of Shor's algorithm, which 
can be described as follows. 
\begin{enumerate}[1.]
\item Given a function $F: G \to S$, create the superposition
\begin{equation*} 
| \Psi \rangle = \frac{1}{\sqrt{|G|}} \sum_{x \in G} |x, F(x) \rangle, 
\end{equation*}
where $G$ is a finite group and $S$ is a set. 
\item Apply a unitary transformation $U : \Psi \mapsto \Psi'$, such as the quantum Fourier transform.
\item Measure $\Psi'$ in the computational basis.
\end{enumerate}

Several quantum algorithms that we will discuss use notions from number theory, algebra and group theory. 
To help the reader's understanding of such algorithms, we will give several brief expositions of these topics. 
This article is based on~\cite{wdc-lec} where the reader can find further details. 

The  article is organized as follows. 
In Section \ref{modular}, we introduce modular arithmetic. 
In Sections \ref{period} and \ref{dlogp} we introduce Shor's algorithm for 
the period finding and the discrete logarithm problem. 
In Section \ref{ahsp} we treat the Abelian hidden subgroup problem and 
introduce some necessary notions to understand the quantum algorithm for the 
Abelian hidden subgroup problem. 
In Section \ref{elliptic}, we describe the elliptic curve problem and its corresponding 
discrete logarithm problem, and 
in Section \ref{unitgp}, we discuss the quantum algorithm for efficiently solving Pell's equation. 
Section \ref{nahsp} introduces the non-Abelian version of the quantum
Fourier transform and discuss the status
of the non-Abelian version of the hidden subgroup problem.

\subsubsection*{Acknowledgements}
These notes are based on a series of lectures by WvD for the 2010 Summer School on Diversities in Quantum Computation/Information at Kinki University, Higashi-Osaka, Japan, which was organized by Mikio Nakahara.   
This material is based upon work supported by the National Science Foundation under Grant No.\,\oldstylenums{0747526}
and by a grant from the Army Research Office with contract number \textsc{w}\oldstylenums{911}\textsc{nf}-\oldstylenums{04}-\textsc{r}-\oldstylenums{0009}.


\section{Modular arithmetic and Residue class of $N$} \label{modular}

We must first introduce a little group theory for a better understanding of 
many quantum algorithms. A group is a combination of a set with a binary operation on its elements that 
obeys certain required group properties. For our purposes here, an important instance  
of a finite group is the residue class of integers modulo $N$ with addition as its group operation. 

Modular arithmetic, $\pmod N$, is understood through a congruence relation on the integers 
${\Z}$, where for integers $a$, $b$ we have  
\begin{equation*}
a \equiv b \pmod N \mbox{\quad if and only if $b-a$ ca be divided by $N$.}
\end{equation*}
\begin{example}
It should be easy to see that $5 \equiv 2 \equiv -1 \pmod{3}$, 
$3 \equiv 15 \pmod{12}$, and $341 \equiv 1 \pmod{10}$.
\end{example} 
Note that modular arithmetic is properly defined for addition, subtraction and multiplication. 
Hence we have for example,  
$2 \times 7 \equiv 1 \pmod{13}$ and 
$(x+y)^2 \equiv x^2 + y^2 \pmod{2}$. 

Modulo $N$ arithmetic classifies the integers ${\Z}$ into $N$ different equivalence classes, which can be indicated 
by the residues of integers when divided by $N$. We thus have  
\begin{equation*} {\Z}/N{\Z} := \{ 0, 1, \dots, N-1 \} \end{equation*}
where $0$ represents $N{\Z}=\{\dots,-N,0,N,2N,\dots\}$, 
the value $1$ represents $1+N{\Z} = \{\dots,-N+1,1,N+1,2N+1,\dots\}$, and so on. 

As mentioned above, modular arithmetic satisfies additiion and multiplication. 
Therefore we can define such arithmetics in ${\Z}/N{\Z}$ by 
\begin{equation*} a + b \pmod{N}, \qquad a \cdot b \pmod{N}. \end{equation*}
Note that for each element $a$ in ${\Z}/N{\Z}$, 
there exists an additive inverse element $b$ (such that $a + b \equiv 0 \pmod{N}$) 
in ${\Z}/N{\Z}$, that is, $N-a$. 
It is straightforward to verify that addition in ${\Z}/N{\Z}$ makes up a finite Abelian group.  What about multiplication in ${\Z}/N{\Z}$? 

Let us compare ${\Z}/5{\Z}$ with ${\Z}/6{\Z}$. 
The number $2$ does not have a multiplicative inverse $\bmod{6}$, 
but it does for $\bmod{5}$. 
It can be shown that a number $y$ has a multiplicative inverse $z$ 
(such that $yz \equiv 1 \pmod N$) 
if and only if the greatest common divisor of $y$ and $N$ obeys $\gcd(y, N) = 1$. 
If $\gcd (y, N) > 1$, then there exists a nonzero $z$ such that 
$yz \equiv 0 \mod N$. 
For instances, $2 \times 3 \equiv 1 \pmod 5$ and $2 \times 3 \equiv 0 \pmod 6$. 
Hence, we see that the set 
\begin{equation*} \left( {\Z}/N{\Z} \right)^{\times} := \{ a \in {\Z}/N{\Z} \ | \ 
\gcd (a,N)=1 \} \end{equation*} 
forms a group with respect to multiplication, which is called 
the \emph{modulo $N$ multiplicative group}.

\section{Period Finding Implies Factoring} \label{period}
Perhaps the best-known application of quantum computers
is its efficient solution to the problem of factoring integers. To explain a closely related algorithm 
for factoring integers due to Miller, we first introduce the period finding problem for the sequence 
$x^0,x^1,x^2,\dots\bmod{N}$. 

For any $x \in ({\Z}/N{\Z})^{\times}$, 
we have an $r$-periodic sequence $x^0=1, x^1, x^2, \dots, x^r=1, x, \dots$. 
The period $r$ is called the \emph{multiplicative order} of $x$ in ${\Z}/N{\Z}$ 
and it is a divisor of the Euler's totient value of $N$
\begin{equation*} 
\phi(N) := |({\Z}/N{\Z})^{\times}| 
= N \prod_{\substack{p|N\\p : \text{prime}}} \left( 1-\frac{1}{p} \right), 
\end{equation*}
which expresses the size of the multiplicative group modulo $N$.
As an example, for $4 \in ({\Z}/9{\Z})^{\times}$, we have 
a sequence $1 =4^0, 4 = 4^1, 7 \equiv 4^2, 1 \equiv 4^3,   4 \equiv 4^4, \dots \pmod{9}$. 
Therefore, the period of $4$ in $({\Z}/9{\Z})^{\times}$ is $3$, a divisor of 
$\phi (9) = 9(1-1/3) = 6$. 

\subsection{Factoring $N$ --- Miller's algorithm}
\begin{theorem}[Miller, 1976]
For a given odd integer $N$ with at least two distinct prime factors, 
we can determine a nontrivial factor of $N$ as follows: 
\begin{enumerate}[1.]
\item Pick up a random $a \in \{ 2,3, \dots, N-1 \}$.
\item Compute $\gcd (a,N)$. If the result is different from 1, 
then it is a nontrivial factor of $N$, and we are done. 
More likely, $\gcd (a,N) = 1$, and we continue.
\item Using a period finding algorithm, 
determine the order $r$ of $a$ modulo $N$.
If $r$ is odd, the algorithm has failed, and we return to
step 1. If $r$ is even, we continue.
\item Compute $\gcd (a^{r/2}-1,N)$. 
If the result is different from 1, then it is a nontrivial factor of $N$. 
Otherwise, return to step 1. 
\end{enumerate}
By repeating this routine we can find all factors of $N$. 
\end{theorem}
With the following lemma, Miller showed that if Step 3 can be executed efficiently, then the above algorithm as a whole is efficient. 
\begin{lemma}
Suppose a is chosen uniformly at random from 
$({\Z}/N{\Z})^{\times}$, where $N$ is an odd integer 
with at least two distinct prime factors. 
Then with probability at least $1/2$, the multiplicative
order $r$ of a modulo $N$ is even, and $a^{r/2} \not\equiv -1 \pmod{N}$.
\end{lemma}

\begin{exercise}[Factoring $N = 21$] 
For all $a \in {\Z}/21{\Z}$ 
figure out what the sequence $a^0, a^1, \dots$ tells us about the
factors of $N = 21$. 
Which $a$ have $\gcd (a, 21) =1$, 
What are the periods? 
Which $a$ give us useful information?
\end{exercise}

\begin{answer}
A modulo multiplication group of ${\Z}/N{\Z}$ equals 
$({\Z}/21{\Z})^{\times} := \{ a \in 
{\Z}/21{\Z} \ | \ \gcd(a, 21) = 1 \} = \{ 1, 2, 4, 5, 8, 10, 11, 13, 16, 17, 19, 20 \}$. 
The following table gives, for all $a \in ({\Z}/21{\Z})^{\times}$, the sequences of values $a^j$, 
the periods $r$ of these sequences, and whether or not this $r$ is useful in determining a nontrivial factor 
of $21$:  
\begin{equation*}
\begin{array}{c|ccccccc|c|c} 
a \setminus j & 0 & 1 & 2 & 3 & 4 & 5 & 6 & \mbox{period $r$} & \mbox{useful?} \\ \hline
1 & 1 & 1 & 1 & 1 & 1 & 1 & 1 & 1 & N \\ 
2 & 1 & 2 & 4 & 8 & 16 & 11 & 1 & 6 & Y\\ 
4 & 1 & 4 & 16 & 1 & 4 & 16 & 1 & 3  & N\\ 
5 & 1 & 5 & 4 & 20 & 16 & 17 & 1 & 6 & N\\ 
8 & 1 & 8 & 1 & 8 & 1 & 8 & 1 & 2 & Y\\ 
10 & 1 & 10 & 16 & 13 & 4 & 19 & 1 & 6 & Y\\ 
11 & 1 & 11 & 16 & 8 & 4 & 2 & 1 & 6 & Y\\ 
13 & 1 & 13 & 1 & 13 & 1 &13 & 1 & 2 & Y\\ 
16 & 1 & 16 & 4 & 1 & 16 & 4 & 1 & 3 & N\\ 
17 & 1 & 17 & 16 & 20 & 4 & 5 & 1 & 6 & N\\ 
19 & 1 & 19 & 4 & 13 & 16 & 10 & 1 & 6 & Y\\ 
20 & 1 & 20 & 1 & 20 & 1 & 20 & 1 & 2 & N
\end{array}
\end{equation*}
\end{answer}

\subsection{Shor's Algorithm}
Shor (1994) proved that period finding can be done
efficiently with a quantum algorithm. As part of this proof, 
he had to show how to efficiently implement the \emph{Quantum Fourier Transform (QFT)} 
over ${\Z}/N{\Z}$: 
\begin{equation*} |x \rangle \mapsto \frac{1}{\sqrt{N}} \sum_{y \in {\Z}/N{\Z}} 
\e^{2\pi \i xy/N} |y \rangle. \end{equation*}
Here is a sketch of Shor's algorithm. 

\begin{algorithm}[Period Finding] \label{alg1}
Let $f:{\Z}/N{\Z} \to S$ 
an $r$-periodic function with
$f(x)=f(y)$ if and only if $(x-y)/r \in {\Z}$ and $r|N$.  
\begin{enumerate}[$(1)$]
\item Create the superposition
\begin{equation*}\frac{1}{\sqrt{N}} \sum_{x \in {\Z}/N{\Z}} |x, f(x) \rangle. \end{equation*}
\item By tracing out the right register, the left register is equivalent to
\begin{equation*} \frac{\sqrt{r}}{\sqrt{N}} \sum_{j \in  \{ 0, 1, \dots , N/r-1\}} |s+jr \rangle \end{equation*}
for a random and unknown $s\in\{0,\dots,r-1\}$.
\item Apply the QFT over ${\Z}/N{\Z}$ to the quantum state, yielding
\begin{equation*} \frac{1}{\sqrt{r}}\sum_{k \in \{ 0, 1, \dots, r-1 \}} \e^{2\pi \i sk/r} | kN/r \rangle. \end{equation*}
\item Sampling the state will gives us $kN/r$ for some random and unknown $k\in\{0,\dots,r-1\}$.
\item By repeating the above procedure several times, a number of multiples of $N/r$ will be obtained. By taking the 
$\gcd$ of these outcomes we learn, with high probability, the value $N/r$ and hence $r$ itself.
\end{enumerate}
\end{algorithm}
Combining the quantum algorithm for efficiently finding periods with Miller's algorithm thus gives us an efficient algorithm for factoring integers.  

\section{Discrete Logarithm Problem} \label{dlogp}
In Section \ref{period} we looked at the problem of determining the period for a 
given $a$ in $({\Z}/N{\Z})^{\times}$ of the sequence $a^0=1, a, a^2, a^3, \dots, a^{r-1}, a^r=1, \dots$. 
Here we consider the problem of finding $\ell$ such that $a^\ell \equiv x \mod N$ for given 
$a$ and $x$ in $({\Z}/N{\Z})^{\times}$. 
The \emph{discrete logarithm} of $x$ with respect to $a$, denoted $\dlog_a (x)$, 
is the smallest non-negative integer $\ell$ 
such that $a^\ell \equiv x \pmod N$. 

This $\dlog$ problem can be generalized in the following way. 
Let $C = \langle g \rangle = \{ g^0=1, g, g^2, \dots \}$ be a 
cyclic group generated by $g$. Then, the discrete logarithm base $g$ of $x\in C$ is denoted 
by $\dlog_g(x)$ and it is again the smallest non-negative integer $\ell$ such that $g^\ell = x$. 
The \emph{discrete logarithm problem} is the problem
of calculating $\dlog_g x$ for a given $x\in C=\langle g\rangle$.

\subsection{Shor's algorithm for calculating discrete logarithms}
Although the problem appears to be difficult for classical
computers, quantum computers can calculate discrete logarithms
efficiently. 
We describe Shor's algorithm~\cite{sh} for discrete logarithm below. 
For simplicity, we assume that the order of the group $N :=|C|$ is known. 
In fact, we can determine it efficiently 
using Shor's algorithm for period finding over ${\Z}$ 
(see \cite[\S~IV.D.]{wdc-lec}). 

\begin{algorithm}[Discrete logarithm] \label{alg2} \
\begin{enumerate}[$(1)$]
\item Create the uniform superposition
\begin{equation*}|{\Z}/N{\Z}, {\Z}/N{\Z} \rangle = 
\frac{1}{N} \sum_{a, b \in {\Z}/N{\Z}} | a, b \rangle. \end{equation*}
\item Define $f : {\Z}/N{\Z} \times {\Z}/N{\Z} \to C$ as follows:
\begin{equation*} f(a, b) = x^{a} g^{b}. \end{equation*}
Note that $f(a,b) = f(c,d)$ if and only if $(a-c) \dlog_g x = (d-b)$. 
Compute this function in an ancilla register, giving
\begin{equation*} \frac{1}{N} \sum_{a,b\in {\Z}/N{\Z}} |a, b, f(a,b) \rangle. \end{equation*}
\item Discard the ancilla register, giving the state 
\begin{equation*} \frac{1}{\sqrt{N}} \sum_{a \in {\Z}/N{\Z}} | a, c-a\dlog_g x \rangle, \end{equation*}
for an unknown $c$. 
\item Perform a Quantum Fourier Transform over ${\Z}/N{\Z} \times 
{\Z}/N{\Z}$ to the two registers and measure its state, 
which will yield the outcome $(\gamma \dlog_g x,\gamma)$ for an arbitrary $\gamma\in{\Z}/N{\Z}$.  
\item Repeat the above steps, thus obtaining another pair of values $(\gamma' \dlog_g x,\gamma')$. With probability at 
least $6/\pi^2 \approx 0.61$ the factors $\gamma$ and $\gamma'$ will be co-prime, hence with the same probability 
the value $\gcd(\gamma \dlog_g x,\gamma' \dlog_g x)$ will equal the desired outcome $\dlog_g x$. 
\end{enumerate}
\end{algorithm}

\subsection{Cryptographic Consequences}
Being able to calculate discrete logarithms
over ${\Z}/N{\Z}$ implies being able to break the Diffie-Hellman 
key-exchange protocol (as well as the ElGamal protocol). 
Unlike our quantum algorithms, the best known 
classical algorithm (the Number Field Sieve) has, in both 
cases, a proven running time of $2^{O(\sqrt{\log N \log \log N})}$ 
and a conjectured running time of $2^{O(\log^{1/3} N (\log \log N)^{2/3})}$. 

Note that the bounds on the classical algorithm are
upper bounds, not lower bounds. 
To find a method of proving that there is no efficient
classical algorithm for factoring or discrete logarithms, is one of the major 
unresolved challenges in computational number theory.

Factoring and the discrete logarithm problem are so-called \emph{natural problems,} 
where the problem statement contains complete information about the problem ($N$ or $N, g, t$). 
In contrast, \emph{black-box problems} have part of the computational problem hidden in a black-box 
that must be queried to find a solution. In this setting, it was proven by Cleve~\cite{cl04} that the 
classical lower bound for the period finding problem is $\Omega (N^{1/3} / \sqrt{\log N})$, 
while quantum mechanically we can solve the same problem with $(\log N)^{O(1)}$ quantum queries.

\section{Abelian Hidden Subgroup Problem} \label{ahsp}
Algorithms \ref{alg1} and \ref{alg2} solve particular instances of a more general problem, 
the \emph{Abelian hidden subgroup problem} (Abelian \HSP). 
Here we will describe this problem and its efficient quantum solution in its generality. 

Let be $G$ a finite Abelian group 
and consider a function $F : G \to S$, where $S$ is some finite set. 
We say that $F$ \emph{hides} the subgroup $H \leq G$ if for all $x,y\in G$ we have 
\begin{align*}
F(x) = F(y)\mbox{~if and only if~} x-y \in H.
\end{align*} 
In other words, $F(x) = F(h+x)$ holds if and only if $h \in H$. 
To understand this situation well, it is helpful to look at the cosets $H_r$ of $H$ in $G$, 
which are defined by $H_r := r+H = \{r+h~:~h\in H\}$.  The \emph{coset decomposition} of $G$ 
in terms of $H$ refers to a mutually disjoint set of cosets $\{H_r~:~r\}$ such that 
$\bigsqcup_r H_r = G$. 
\begin{example}
Take $G = {\Z}/6{\Z} := \{ 0, 1, 2, 3, 4, 5 \}$ and $H=\{ 0, 3 \} \leq G$. 
Then we see that $G$ can be decomposed  as 
$G= H_0 \sqcup H_1 \sqcup H_2$, 
where $H_0 := 0+H = \{ 0, 3 \}$, $H_1:= 1+H = \{ 1, 4 \}$, and $H_2:= 2+H = \{ 2, 5 \}$ are mutually disjoint 
cosets of $H$.  
\end{example}

Going back to the Hidden Subgroup Problem, we see that a function $F$ that hides $H$ 
is constant function on each coset $r+H$ and injective 
on the different cosets $r+H$. 
In the Abelian \HSP, we are asked to find a generating set for $H$ 
given the ability to query the function $F$. 
For every different kind of group $G$ and its possible
subgroups $H$, this is a different kind of problem. 

\subsection{Character and Dual group}
To explain the efficient quantum solution to the Abelian \HSP, we have to introduce the notion of 
a character over the finite Abelian group. 
For a finite Abelian group $G$, a \emph{character} over $G$ is a function 
$\Psi:G \to {\C}^{\times} := \{ z \in {\C} \ | \ |z| = 1 \}$ 
with the property that for all $x,y\in G$ we have $\Psi (x+y) = \Psi(x) \Psi(y)$. 
The set of all possible characters over $G$ is denoted by 
\begin{equation*} 
\hat{G} := \{ \Psi:G\rightarrow \C \ | \text{$\Psi$ is a character over $G$} \},  \end{equation*}
called the \emph{dual group} of $G$. 
The \emph{trivial character} of $G$ is the unit function  $\Psi (x) = 1$. 

\begin{example}
In the case of $G={\Z}/N{\Z}$ we have the characters $\Psi_a (x) = \e^{2\pi \i ax/N}$
for $a \in \{ 0,1, \dots, N-1 \}$ with $a=0$ yielding the trivial $\Psi = \id$. 
We can also define a group operation $\circ$ on $\hat{G}$ by $\Psi_a \circ \Psi_{a+b}$, 
showing that $(\hat{G},\circ)$ is isomorphic to $\Z/N\Z=G$, and hence $|G|=|\hat{G}|$.  
\end{example}

The above isomorphism between $G$ and $\hat{G}$ is no coincidence, as for all finite Abelian groups 
we have $G\simeq \hat{G}$. As another example, consider a finite cyclic group $G := \{ 1, g, g^2, \dots, g^{r-1} \}$ of size $r$ 
generated by $g$. Now the $r$ different characters of $G$ are the functions 
$\Psi_a$ defined by $\Psi_a(g^k) = \e^{2\pi\i ak/r}$ for all $a,k\{0,\dots,r-1\}$. 
Again the dual group $\hat{G}$ forms a group with respect to the composite of mapping 
$(\Psi_a \circ \Psi_b (x) = \Psi_a (x) \cdot \Psi_b (x) = \Psi_{a+b}(x)$ for $x \in G$. 
As $G\simeq \Z/r\Z$ we have again $G \simeq \hat{G} \simeq \Z/r\Z$ and consequently $|\hat{G}| = |G|$. 

Characters have several useful properties, the following which 
are fundamental tools to describe quantum algorithm for the finite Abelian \HSP. 
\begin{lemma} \label{lem-ch}
Let $G$ be a finite Abelian group and $H$ a subgroup of $G$. 
For each character $\Psi$ on $G$,  we have 
\begin{equation*}
\frac{1}{|G|} \sum_{x \in G} \Psi (x) = 
\begin{cases}
1 & \text{if $\Psi$ is trivial} \\
0 & \text{if $\Psi$ is nontrivial}
\end{cases}
\end{equation*}
More specifically we have 
\begin{equation*}
\frac{1}{|H|} \sum_{x \in H} \Psi (x) = 
\begin{cases}
1 & \text{if $\Psi$ is trivial on $H$} \\
0 & \text{if $\Psi$ is nontrivial on $H$.}
\end{cases}
\end{equation*}
\end{lemma}

\subsection{Quantum Fourier transform for Abelian Groups}
In Section \ref{period}, we have described the quantum Fourier transform over 
${\Z}/N{\Z}$. This transformation can be generalized to any finite Abelian group 
by using the characters over $G$. 
For each finite Abelian group $G$, we define the unitary \emph{quantum Fourier transform} $\C^{|G|}\rightarrow \C^{|G|}$ 
over $G$ by
\begin{equation*} | x \rangle \mapsto \frac{1}{\sqrt{|\hat{G}|}} \sum_{\Psi \in \hat{G}} \Psi (x) | \Psi \rangle. \end{equation*}
for each $x \in G$. 
If we know enough about $G$, we can 
implement this efficiently in $\poly (\log |G|)$ steps.
The quantum Fourier transform will help us solve the general Abelian hidden subgroup problem.

\subsection{Solving Abelian Hidden Subgroup Problem}
\begin{algorithm}
Let $G$ be a finite Abelian group and let be $F:G \to S$ be a function that hides a subgroup $H \leq G$.
The following efficient quantum algorithms determines (a set of generators of) $H$.  
\begin{enumerate}[$(1)$]
\item Create the superposition: 
\begin{equation*} \frac{1}{\sqrt{|G|}}\sum_{x \in G} |x, F(x) \rangle \end{equation*}
When ignoring the right register and since $F$ hides $H$, the left register can be described by 
\begin{equation*} |s+H \rangle := \frac{1}{\sqrt{|H|}} \sum_{x \in H} |s+x \rangle\end{equation*} 
for an unknown $s$. 
\item Apply the quantum Fourier transform over $G$ to the left register, 
\begin{align*}
\ket{s+H} & \mapsto \frac{1}{\sqrt{|H| \cdot |G|}} \sum_{\Psi \in \hat{G}} \left( \sum_{x \in H} \Psi (s+x) \right) | \Psi \rangle \\
& = \sqrt{\frac{|H|}{|G|}} 
\sum_{\Psi \in \hat{G}} \Psi (s) \left( \frac{1}{|H|} \sum_{x \in H} \Psi (x) \right) |\Psi \rangle 
\end{align*}
\item Note that only the $\Psi$s that are trivial on $H$ survive
the summation $x \in H$. Hence when measuring the register, we will only observe such $\Psi$s 
of which there are $|G|/|H|$. 
\item By repeating the above procedure $(\log |G|)^{O(1)}$ times, we obtain enough information, 
through the observed characters that are trivial on $H$ to reconstruct $H$. 
\end{enumerate}
\end{algorithm}

\subsection{Hidden Periodicity Problem over ${\Z}$} \label{hiddenz}
In the previous section, we saw how the Abelian hidden subgroup problem can 
be solved efficiently over any known finite Abelian group. An important generalization 
of this problem is the \emph{hidden periodicity problem} over $\Z$, 
where $F$ is a function defined over ${\Z}$ with a period $p$, i.e.\ 
 $F(x) = F(y)$ if and only if $x \equiv y \mod p$. With $G$ the infinite Abelian group $\Z$ and 
$H=p\Z$ its subgroup, this is yet another instance of the Hidden Subgroup Problem, but this time
for infinite groups. 
As explained, for example, in \cite[\S IV.D]{wdc-lec} 
Shor~\cite{sh} showed how to find the period $p$ hidden by $F$ 
efficiently in time $\poly (\log p)$ on a quantum computer.

\subsection{Decomposing Abelian Groups} \label{sec:groupstructure}
It is known that any finite Abelian group $(G,+)$ has a decomposition 
\begin{equation*}
G \simeq {\Z}/p_1^{r_1}{\Z} \oplus \cdots \oplus 
{\Z}/p_k^{r_k}{\Z}. 
\end{equation*}
Suppose we are given implicitly the encoding $E:G \to S$ of a
finite Abelian group $(G,+)$, with the properties:
\begin{enumerate}
\item $E(g)$ is an injection (i.e. $E(h) = E(g)$ implies $g=h$).
\item Given $E(g)$ and $E(h)$ you can compute $E(g+h)$ and $E(-g)$. 
\end{enumerate}
Then there exists a quantum algorithm to solve the encode $E$ efficiently. 
Further, we can find the above decomposition by this algorithm. 
Watrous~\cite{wa01} generalized this quantum algorithm to all ``solvable'' finite groups.

\section{Fields} \label{field}
In this section we turn out attention to quantum algorithms that deal with \emph{fields} instead of groups.  
A typical example of a field is given by the set of rational numbers ${\Q}$, which is closed for addition and multiplication, 
and each rational number $x$ has an additive inverse number $y$ and 
a multiplicative inverse number $z$ such that $x+y = 0$ and $x \cdot z = 1$. 
In general a field $F$  is a set that is closed under an addition ($+$) and a multiplication ($\cdot$)
operation 
\begin{alignat*}{2}
+ : & \ F \times F \to F, &\qquad  (x,y) &\mapsto x+y, \\
\cdot : & \ F \times F \to F, & \qquad (x,y) &\mapsto x \cdot y
\end{alignat*}
and that satisfies that any $a, b, c, u, v \in F$, 
\begin{enumerate}
\item $a +(b+c) = (a+b)+c, \quad a \cdot (b\cdot c) = (a \cdot b) \cdot c$ (Associative), 
\item $a+b = b+a, \quad  a \cdot b = b \cdot a$ (Commutative),
\item $u \cdot (a+b) = u \cdot a + u \cdot b, \quad 
(u+v) \cdot a = u \cdot a + v \cdot b$ (Distributive)
\item There exist elements $0$ and $1$ such that $a + 0 = 0 + a = a$ 
and $a \cdot 1 = 1 \cdot a$ for all $a \in F$. 
Such elements are unique and called the zero element and the unit element, 
respectively. 
\item For each $a$ in $F$, there exists an element $b$ in $F$ such that 
$a + b = b + a = 0$. This element is called an additional inverse element of $a$. 
\item For each $a$ in $F \setminus \{ 0 \}$, there exists an element $c$ in $F$ such that 
$a \cdot c = c \cdot a = 1$. 
This element is called a multiplicative inverse element of $a$. 
\end{enumerate}
Unsurprisingly, when the number of elements in $F$ is finite, $F$ is called a \emph{finite field.} 
Given a finite size $q=|F|$ there can essentially be only one field with that size, and 
we denote this finite field by ${\F}_q$. 
It is known that when $F$ is a finite field, it must hold that $|F| = p^n$ with $p$ a prime integer, and $n\in\N$.   

\begin{example} The following examples and counterexamples of fields are standard. 
\begin{enumerate}[1.]
\item ${\Q}, {\R}, {\C}$ are  fields. However 
${\Z}$ is not a field as not all integers have a multiplicative inverse in $\Z$. 
\item For any prime number $p$, ${\Z}/p{\Z}$ is a finite field and we often write ${\F}_p$ instead of ${\Z}/p{\Z}$.  
\item For $N$ a composite number, ${\Z}/N{\Z}$ is not a finite field. 
\end{enumerate}
\end{example}

\begin{exercise}
Let be ${\F}_4 = \{ 0, 1, x, y \}$. Write down its addition and multiplication tables. 
\end{exercise}
\begin{answer}
We start with the multiplication table for all $a,b\in{\F}_4$: 
\begin{equation*}
\begin{array}{c|cccc} 
a \cdot b & b=0& b=1 & b=x & b=y \\ \hline
a=0& 0& 0& 0& 0 \\
a= 1 & 0& 1 & x & y \\
a= x & 0 & x & x^2 & xy \\
a= y & 0 & y & yx & y^2 
\end{array}
\end{equation*}
Our task is to determine the values on the bottom right $2 \times 2$ block in the above table. 
If $xy =y$, we have $x = 1$ since $y$ has an inverse element, which implies a contradiction. 
Similarly $xy$ is not equal to $x$ and hence we have $xy = 1$. 
As this implies that $yx = 1$ we see that $x^2$ and $y^2$ 
must be $y$ and $x$, respectively. Hence the following multiplication table for $a,b\in\F_4$ is the correct one. 
\begin{equation*}
\begin{array}{c|cccc} 
a\cdot b & b=0 & b=1 & b=x & b=y \\ \hline
a=0 & 0 & 0 & 0 & 0 \\
a=1 & 0 & 1 & x & y \\
a=x & 0  & x & y & 1 \\
a=y & 0  & y & 1 & x 
\end{array}
\end{equation*}
What remains is to determine the rules for addition in $\F_4$. As $\F_4$ is an extension field of $\F_2$ we have $1+1=2\equiv 0 \pmod{2}$, which gives the following table for $a+b$ for all $a,b\in\F_4$ 
\begin{equation*}
\begin{array}{c|cccc} 
a+b & b=0 & b=1 & b=x & b=y \\ \hline
a=0 & 0 & 1 & x & y \\
a=1 & 1 & 0 & 1+x & 1+y \\
a=x & x & x+1 & 0 & x+y \\
a=y & y & y+1 & y+x & 0 
\end{array}
\end{equation*} 
As in the case of the multiplicative calculation, $1+x$ is not equal to $0$, $1$ and $x$, and 
hence we have $1+x = x+1= y$, which shows that $x+y =1$ and $1+y = y+1 = x$. 
Hence we end up with this table: 
\begin{equation*}
\begin{array}{c|cccc} 
a+b & b=0 & b=1 & b=x & b=y \\ \hline
a=0 & 0 & 1 & x & y \\
a=1 & 1 & 0 & y & x \\
a=x & x & y & 0 & 1 \\
a=y & y & x & 1 & 0 
\end{array}
\end{equation*} 
\end{answer}

\subsection{Field extensions}
Note that the field  of real numbers ${\R}$ 
is included in the field of complex numbers ${\C}$. 
In general, when a field $K$ contains a field $F$ as a subset, 
$F$ is called a \emph{subfield} of $K$. Conversely, $K$ is called an \emph{extension field} of $F$. 
In this example, the extension field $K$ of $F$ can be viewed as a vector space over $F$ and 
its dimension is called the \emph{degree} of $K$ over $F$. 
The complex numbers ${\C}$ is an extension field of ${\R}$ with degree $2$. 
Another way of understanding such algebraic extensions is by viewing ${\C}$ as
$\R$  extended  with the solution of the degree $2$ equation $X^2 + 1 = 0$, 
which allows us to write $\C = \R(X^2+1)$.  
Similarly, the finite field ${\F}_4$ is a degree $2$ extension of $\F_2$ with an $X$ such that $X^2+X=1$. 
Continuing with this idea, ${\F}_4$ can be further extended to ${\F}_{16}$ and in general, ${\F}_s$ is an extension of ${\F}_q$ if and only if $s$ is an integral power of $q$. 

\subsection{Number Fields}
A complex number $\alpha$ is an \emph{algebraic number} if $\alpha$ satisfies some 
monic polynomial with rational coefficients, 
\begin{equation*} p(\alpha) = 0, \qquad p(x) = x^n + c_1 x^{n-1} + \cdots + c_{n-1} x + c_n = 0 
\quad (c_j \in {\Q}). \end{equation*}
In particular, $\alpha$ is an \emph{algebraic integer} if $\alpha$ satisfies some 
monic polynomial with integral coefficients. 
Any (usual) integer $z \in {\Z}$ is an algebraic integer, 
since it is the zero of the linear monic polynomial $p(x) = x - z$. 

A \emph{number field} is an extension of ${\Q}$ with some 
algebraic numbers. For instance, 
\begin{equation*} K= {\Q} (\sqrt{-5}) := \{ a+b\sqrt{-5} \ | \ a, b \in {\Q} \} \end{equation*}
is a \emph{quadratic} number field of degree $2$, 
since $\sqrt{-5}$ is a root of the equation $x^2+5=0$. 
The set of algebraic integers contained in a number field $K$, 
is called the \emph{ring of integers} $\mathcal{O}_K$ of $K$. 
One can show that $\mathcal{O}_K$ is a ring. 

For example, the ring of integers of ${\Q}$ is ${\Z}$ and that of 
${\Q} (\sqrt{m})$ ($m$ is a square-free integer) is 
\begin{equation*}  \mathcal{O}_{{\Q} (\sqrt{m})}
 = \{ a+b \omega \ | \ a, b \in {\Z} \}, \end{equation*}
where 
\begin{equation*}
\omega = \begin{cases}
\sqrt{m} & \text{if $m \equiv 2, 3 \pmod{4}$}, \\
\frac{1+\sqrt{m}}{2} & \text{if $m \equiv 1 \pmod{4}$}. 
\end{cases}
\end{equation*}
Note that, unlike $\Z$, a ring of integers of $K$ in general does not satisfy the unique factorization property.  
For example, $6 = 3 \times 2 = (1+\sqrt{-5})(1-\sqrt{-5})$ 
in $\mathcal{O}_{{\Q}(\sqrt{-5})}$.

\section{Elliptic Curve Cryptography} \label{elliptic}
\subsection{Elliptic Curves}
Let $K$ be a field and consider the cubic equation $Y^2 = X^3+ aX^2 + bX + c$ with $a,b,c \in K$. 
If this equation is nonsingular, the corresponding \emph{elliptic curve} $E(K)$ is the set 
of its solutions $(X,Y) \in K^2$ combined with ``the point at infinity'' $\mathcal{O}$: 
\begin{equation*} E(K) := \{ (X,Y) \in K^2 \ | \ Y^2 = X^3+ aX^2 + bX + c \} \cup \{ \mathcal{O} \}. \end{equation*}
By suitable linear transformations, any elliptic curve can be rewritten 
in the form of the Weierstra{\ss} equation 
\begin{equation*} Y^2 = X^3 + \alpha X + \beta \qquad (\alpha, \beta \in K). \end{equation*}

\begin{exercise}
Consider an elliptic curve defined by $Y^2 = X^3 + 2X +1$ over ${\F}_5$. 
List the solutions $(X, Y) \in {\F}_5^2$. 
\end{exercise}

\begin{answer}
\begin{align*}
E({\F}_5) :=& \{ (X,Y) \ | \ Y^2 = X^3 + 2X +1 \} \cup \{ \mathcal{O} \} \\
=&\{ (0,1), (0,4), (1,2), (1,3), (3,2), (3,3), \mathcal{O} \}. 
\end{align*}
\end{answer}
Surprisingly, for the elements of $E(K)$ we can define an addition operation. 
To make the definition of addition easier to understand, we will consider elliptic curves 
over ${\R}$ for the moment. 
Given two points $P, Q \in E$, their sum $P+Q$ is defined geometrically as follows. 
First assume that neither point is $\mathcal{O}$. 
Draw a line through the points $P$ and $Q$ or, if $P=Q$ draw the 
tangent to the curve at $P$ and let denote $R$ the third 
point of intersection with $E(K)$ (if the line is parallel to $X=0$, we have the intersection $R=\mathcal{O}$). 
Then we define $P+Q$ by the reflection of $R$ about the $x$ axis,
where the reflection of $\mathcal{O}$ is itself. If one of $P$ or $Q$ is $\mathcal{O}$, 
we draw a vertical line through the other point, so that $P+\mathcal{O}=P$, showing that $\mathcal{O}$ is the zero element. 
Reflection about the $X$ axis corresponds to negation, so we can think of the rule as saying 
that the three points of intersection of a line with $E(K)$  sum to $\mathcal{O}$. 

The above geometrical introduction does not necessarily make sense for other fields. 
Nevertheless, we can take the same structure for any field $K$ 
by translating the above argument in terms of coordinates in $K^2$. 
Let be $P= (x_P, y_P)$ and $Q=(x_Q,y_Q)$. Provided
$x_P \neq x_Q$, the slope of the line through $P$ and $Q$ is 
\begin{equation*} \lambda = \frac{y_Q - y_P}{x_Q - x_P} \end{equation*}
Computing the intersection of this line with the elliptic curve $E$, we find
\begin{align*}
x_{P+Q} &= \lambda^2 - x_P - x_Q,\\
y_{P+Q} &= \lambda(x_P-x_{P+Q})-y_P.
\end{align*}
If $x_P = x_Q$, there are two possibilities for $Q$: either 
$Q = (x_Q, y_Q) = (x_P, y_P)=P$ or $Q=(x_Q, y_Q) = (x_P, -y_P) = -P$. 
If $Q=-P$, then $P+Q=\mathcal{O}$. On the other hand, if $P=Q$, that is, 
if we are computing $2P$, then the two equalities  hold with $\lambda$ 
replaced by the slope of the tangent to the curve at $P$, namely, 
$\lambda = \frac{3x_P^2+a}{2y_P}$, 
unless $y_P=0$, in which case the slope is infinite, so $2P=\mathcal{O}$. 

\subsection{Elliptic Curve Cryptography}
The discrete logarithm problem for an elliptic curve $E$ defined over 
a finite field ${\F}$ is described as follows. Given two points $P$ and $Q$ in $E$, 
how many times $r\in\N$ do
we need to add $P$ to get 
\begin{equation*} rP = \underbrace{P+\cdots+P}_r = Q \ \text{?} \end{equation*}
This problem appears harder than the discrete logarithm over $(\Z/N\Z)^\times$ 
and the best known classical algorithm for this problem has a time complexity 
of $\Omega(\sqrt{|{\F}|})$. 
As a result, elliptic curve cryptography systems that rely on the hardness of the discrete logarithm 
over elliptic curves allow smaller keys.  An example of such a system supported by Certicom and is used in Blackberries.

Our quantum algorithm for solving the Hidden Subgroup Problem over $(E,+)$ still applies however, and thus
allows us to break this crypto-system as well. 
For more details on the implementation of Shor's algorithm over
elliptic curves, see Proos and Zalka~\cite{pz03}, Kaye~\cite{k05},
and Cheung et al.~\cite{cmmp08}

\section{Counting Points of Finite Field Equations}
As in the case of elliptic curve over finite fields, for $f \in {\F}_q [x,y]$ a polynomial 
in two variables with coefficients in ${\F}_q$, the finite set of zeros of $f$ make a curve. 
Our interest lies with the number of zeros, that is, the number of points of the curve 
$C_f := \{ (x,y) \in {\F}_q^2 | f(x,y)=0 \}$.  
A key parameter characterizing the complexity 
of this counting problem is the size $q$ of the 
field $q (=p^r)$ and  the genus $g$ of the curve.  
For a nonsingular, projective, planar curve $f$, the genus is 
$g = (d-1)(d-2)/2$, where $d = \deg (f)$ is the degree of the polynomial. 
Elliptic curves have genus $1$. 

In the case of the classical algorithm, Schoof~\cite{sc85} 
described an algorithm to count the number 
of points on an elliptic curve over ${\F}_q$ 
in time $\poly (\log q)$. Subsequent results by Pila~\cite{p90}, 
Adleman and Huang~\cite{ah01} generalized this result to hyper-elliptic 
curves, giving an algorithm with running time 
$(\log q)^{O(g^2 \log g)}$.  
For fields ${\F}_{p^r}$, Lauder and Wan~\cite{lw02} showed the 
existence of a deterministic algorithm for counting points with 
time complexity $\poly (p, r, g)$. All these classical algorithms are 
bested by the quantum algorithm that Kedlaya~\cite{ke06} developed, 
which solves the same counting problem with time complexity
$\poly (g, \log q)$. 

\subsection{Ingredients of Kedlaya's Algorithm}

Kedlaya's algorithm is based on the relation between the \emph{class group} of a curve $C_f$ and 
the zeros of the Zeta-function of $f$. 
Let $f \in {\F}_q [x,y]$ be a  polynomial and let 
$C_f := \{ \boldsymbol{x} \in {\P}^2 ({\F}_q) \ | f(\boldsymbol{x}) = 0 \} $
be its smooth curve in the projective plane ${\P}^2 ({\F}_q)$. 
The projective plane is defined by 
${\P}^2 ({\F}_q) := \left\{ (x,y,z) \in {\F}_q^{3} 
\setminus \{ \boldsymbol{0} \}\right\}/\equiv$
with the equivalence  $(x,y,z)\equiv (x',y',z')$ if and only if there 
exists an $\alpha\in\F_q$ such that $(x',y',z') = \alpha(x,y,z)$. 
For each exponent $r\in\N$ we define $N_r$ to be the number of zeros of $f$ in $\P^2(\F_{q^r}$:
\begin{equation*} 
N_r := |\{ \boldsymbol{x} \in {\P}^2({\F}_{q^r}) \ | \ f(\boldsymbol{x}) = 0 \}|. 
\end{equation*}
Using these $N_r$, the Zeta-function of a curve $C$ is  defined as the power series
\begin{equation*}
Z_C(T) := \exp\left(
{\sum_{r=1}^\infty\frac{N_rT^r}{r}}. 
\right)
\end{equation*}
It is not hard to see that knowing $Z_C$ implies knowing the values $N_r$. 

Kedlaya's quantum algorithm exploits the fact that the sizes of the class groups of 
$C_f$ will give us information about the function $Z_C(T)$. 
It is known that these class groups are a finite Abelian groups and as was discussed
in Section~\ref{sec:groupstructure}, quantum computers are able to find the structure and size of a given 
finite Abelian group. Through this connection from the size of the class groups, through the properties of the Zeta function, Kedlaya's algorithm determines the values $N_r$ efficiently in terms of the $\log q$ and the 
degree $d$ of the polynomial $f$. 

\section{Unit Group of Number Fields} \label{unitgp}
Let be $K$ a number field and $\mathcal{O}_K$ the ring of integers of $K$. 
The \emph{unit group} of $K$ is the set of all multiplicatively invertible elements of $\mathcal{O}_K$. 
We denote the unit group of $K$ by
\begin{equation*} \mathcal{O}_K^{\times} := \{ u \in \mathcal{O}_K \ | \ \text{$\exists u^{-1} \in \mathcal{O}_K$ 
such that $u \cdot u^{-1} = 1$} \}. \end{equation*}
As an example, let us consider the case of the number field $K={\Q} (\sqrt{5})$ and its
 ring of integers  
\begin{equation*} \mathcal{O}_{{\Q} (\sqrt{5})} = \{ a + b\tfrac{1+\sqrt{5}}{2} \ | \ a, b \in {\Z} \}. \end{equation*}
The element $9+4\sqrt{5}$ is a unit of ${\Q} (\sqrt{5})$ as it has an 
inverse element in the ring of integers: 
$(9+4\sqrt{5})(9-4\sqrt{5}) = 1$. 
Furthermore, all powers $(9\pm 4\sqrt{5})^k$ 
will be units as well. As an aside, note that these units $x\pm y\sqrt{5}$ are exactly the solutions to \emph{Pell's equation}
\begin{equation*} x^2 - m y^2 = 1. \end{equation*}
for $m=5$. 
In general it is known that for $m\in\N$ and the corresponding real quadratic field $K=\Q(\sqrt{m})$
there exists a \emph{fundamental unit} $\varepsilon_0 \in \mathcal{O}_K^{\times}$ such that 
\begin{equation*} \mathcal{O}_K^{\times} = \{ \pm \varepsilon_0^n \ | \ n \in {\Z} \} \end{equation*}

Quantum computers can exploit the fact that an integer $x \in {\Z} [\sqrt{m}]$ is a unit if and only if
\begin{equation*} x {\Z} [\sqrt{m}]  ={\Z} [\sqrt{m}]. \end{equation*} 
Hence, the function $h(z) = \e^z {\Z} [\sqrt{m}]$ is periodic with 
period $\mathcal{R} := \log \varepsilon_0$ where $\varepsilon_0$ is the above mentioned fundamental unit of 
${\Q}(\sqrt{m})$. 
Hallgren~\cite{he07} 
showed that Shor's original period finding algorithm can be extended to find this period $\mathcal{R}$ 
in this number field setting.

\section{Non-Abelian Hidden Subgroup Problems} \label{nahsp}
In the previous sections, we saw that the Abelian Fourier transform 
can be used to exploit the symmetry of an Abelian hidden subgroup problem, 
and that this essentially gave a complete solution. 
We would like to generalize this to non-Abelian groups.

\subsection{Hidden Subgroup Problem}
Let be $G$ a non-Abelian group. We say that a function $F: G \to S$ 
\emph{hides} a subgroup $H \leq G$ if for all $x,y\in G$
\begin{equation*} \text{$F(x) = F(y)$ if and only if $x^{-1} y \in H$.} \end{equation*}
In other words, $F$ is constant on left cosets 
$H, g_1 H, g_2 H, \dots$ of $H$ in $G$, and distinct on different left cosets. 
The non-Abelian hidden subgroup problem is to determine $H$ from $F$. 
For every different kind of group $G$ and its possible
subgroups $H$, this is a different kind of problem. 

\subsection{Example: Graph Automorphism Problem}
As an instance of the non-Abelian hidden subgroup problem, we describe the 
graph automorphism problem. 
A graph is an ordered pair $G = (V, E)$ comprising a set $V := \{ 1, \dots, n \}$ 
of vertices or nodes together with a set $E$ of edges, 
which are $2$-element subsets of $V$ such that $(i, j)\in E\subseteq V^2$ implies that $i$ and $j$ are connected. 

Let us now consider permutations $\pi\in S_n$ of the vertices of $G$, which are defined by $\pi G = G' = (V,E')$ 
with $E' = \{(\pi(i),\pi(j))~|~(i,j)\in E\}$.  
For some permutations we will have $\pi G = G' = G$, for some others we will have $G'\neq G$.  
An \emph{automorphism} of a graph $G = (V,E)$ is a such a permutation with $\pi G = G$. 
The set of automorphisms of $G$ is a subgroup of the symmetric group of degree $S_n$ and we 
denote this automorphism group by $\aut (G)$. 

The graph automorphism problem is the problem of determining all 
automorphisms of a given graph $G$. Using the function $F(\pi)= \pi G$ defined over the symmetric group, 
this is a Hidden Subgroup Problem over $S_n$.  

\subsection{Some Representation Theory}
To describe potential quantum algorithms for the non-Abelian hidden subgroup problem, we have to introduce 
some notions of the representation theory. 

Let be $G$ a finite  group. 
A \emph{representation} of $G$ over the vector space ${\C}^n$ is a function 
$\rho : G \to GL ({\C}^n)$ with the property $\rho (x \cdot y) = \rho (x) \rho (y)$ 
for any $x, y \in G$, where $GL ({\C}^n)$ is the group of 
all invertible, linear transformations of ${\C}^n$, called the general linear group of ${\C}^n$. 
We thus see that $\rho$ is a homomorphism from the group $G$ to the group $GL ({\C}^n)$. 
We can easily show that $\rho (1) = I$, which is the $n$-dimensional unit matrix 
and $\rho (x^{-1}) = \rho (x)^{-1}$. 
We say that ${\C}^n$ is the representation space of $\rho$, 
where $n$ is called its \emph{dimension} (or \emph{degree}), denoted $d_{\rho}$. 
The $1$-dimensional case implies that the representations are the characters mentioned in Section~\ref{ahsp}. 
Note that the general linear group $GL ({\C}^n)$ is also non-abelian for $n \geq 2$. 
For all finite groups, the representation $\rho$ is a unitary representation, that is, 
one for which $\rho (x)^{-1} = \rho (x)^{\dagger}$ for all $x \in G$.  

Given two representations $\rho: G \to V$ and $\rho':G \to V'$, 
we can define their direct sum, a representation $\rho \oplus \rho' : G \to V \oplus V'$ 
of dimension $d_{\rho \oplus \rho'} = d_{\rho} + d_{\rho'}$. 
The representation matrices of $\rho \oplus \rho'$ are of the form
\begin{equation*} (\rho \oplus \rho') (x) = \left(
\begin{tabular}{cc}
$\rho (x)$ & $0$ \\
$0$ & $\rho' (x)$
\end{tabular} \right)
 \end{equation*}
for all $x$ in $G$. 
A representation is \emph{irreducible} if it cannot be decomposed 
as the direct sum of two other representations. 
Any representation of a finite group $G$ can be written as a direct sum of irreducible 
representations of $G$. We denote a complete set of irreducible representations of $G$ 
by $\hat{G}$. 

Another way to combine two representations is the 
tensor product. The tensor product of $\rho : G \to V$ and 
$\rho' :G\to V'$ is $\rho \otimes \rho': G\to V \otimes V'$, 
a representation of dimension $d_{\rho \otimes \rho'} = d_{\rho} d_{\rho'}$. 

\begin{exercise}
Consider the non-Abelian group $G = \langle A, B \rangle$ 
generated, under multiplication, by the following two matrices 
$A$ and $B$: 
\begin{equation*} A=\left( 
\begin{tabular}{ccc}
0 & 1 & 0 \\
0 & 0 & 1 \\
1 & 0 & 0 
\end{tabular} \right), \quad 
B=\left( 
\begin{tabular}{ccc}
1 & 0 & 0 \\
0 & 0 & 1 \\
0 & 1 & 0 
\end{tabular} \right). 
\end{equation*} 
How many elements does this group have?
A representation is called \emph{faithful} if for all $x \neq y$ 
we have $\rho (x) \neq \rho (y)$, 
i.e.\ when $\rho$ is injective. 
Find a faithful representation with $d=2$ for $G \langle A, B \rangle$.  
\end{exercise}

\begin{answer}
Observe the left-action and 
right-action of $A$: 
\begin{equation*}
AB = \left( 
\begin{tabular}{ccc}
0 & 0 & 1 \\
0 & 1 & 0 \\
1 & 0 & 0 
\end{tabular} 
\right), \ 
BA = \left( 
\begin{tabular}{ccc}
0 & 1 & 0 \\
1 & 0 & 0 \\
0 & 0 & 1 
\end{tabular} 
\right). \end{equation*}
Hence, $ABA=B$. Note that 
$A^3 = I_3$ and $B^2 = I_3$. 
Therefore, any element of $\langle A, B \rangle$ can be expressed as 
$A^l B^k$ with $l\in\{0,1,2\}$ and $k\in\{0, 1\}$, which means that $|\langle A, B \rangle| = 6$. 

The second problem is how to find a representation $\rho$ over $\langle A, B \rangle$. 
For this purpose, we exploit the algebraic properties $A^3 = I_3$ and $B^2 = I_3$. 
Since $\rho$ satisfies $\rho (X Y) = \rho (X) \rho (Y)$ and $\rho (I_3) = I_2$, 
the images of $A$ and $B$ for $\rho$ must satisfy $\rho (A)^3 = I_2$ and $\rho (B)^2 = I_2$. 
Therefore, this problem is equivalent to finding elements $\rho(A)=S$ and $\rho(B)=T$ 
in $GL_2 ({\C})$, such that $S^3 = I_2$, $T^2 = I_2$
and all $S^lT^k$ are different. 
These requirements are met, for example, by the following two matrices
\begin{align*}
S= \left( 
\begin{array}{cc}
\e^{2\pi\i/3} & 0 \\
0 & \e^{-2\pi\i/3} 
\end{array} \right) 
\quad \text{and} \quad 
T = \left( 
\begin{array}{cc}
0 & 1 \\
1 & 0 
\end{array}
\right). 
\end{align*}
\end{answer}

\subsection{Non-Abelian Fourier transform}
For a finite non-Abelian group $G$, we define the $|G|$ dimensional 
quantum Fourier transform over $G$ for every $x\in G$ by
\begin{equation*}
 |x \rangle \mapsto \frac{1}{\sqrt{|G|}} \sum_{\rho \in \hat{G}} d_{\rho} 
|\rho, \rho (x) \rangle, 
\end{equation*}
 where $|\rho \rangle$ is a state 
that labels the irreducible representations, 
and $|\rho (x) \rangle$ is a normalized $d_{\rho}^2$-dimensional state whose 
amplitudes are given by the matrix entries of the $d_{\rho} \times d_{\rho}$ matrix $\rho(x)$: 
\begin{align*}
| \rho (x) \rangle :=& (\rho (x) \otimes I_{d_{\rho}}) \sum_{j=1}^{d_{\rho}} 
\frac{|j,j \rangle}{\sqrt{d_{\rho}}} 
= \sum_{j,k=1}^{d_{\rho}} \frac{\rho (x)_{j,k} |j,k \rangle}{\sqrt{d_{\rho}}}.
\end{align*}
It can be shown that this quantum Fourier transform over $G$ is a unitary matrix 
\begin{align*}
\sum_{x \in G} | \hat{x} \rangle \langle x | 
= \sum_{x \in G} \sum_{\rho \in \hat{G}} 
\sqrt{\frac{d_{\rho}}{|G|}} 
\sum_{j,k=1}^{d_{\rho}} \rho (x)_{j,k} | \rho, j,k \rangle \langle x |. 
\end{align*}
Note that the Fourier transform over a non-Abelian $G$ is not 
uniquely defined, rather, it depends on a choice of basis for 
each irreducible representation of dimension greater than $1$.

\subsection{Fourier Sampling}
Applying the Fourier transformation to a superposition, we obtain 
\begin{equation*} \sum_{x\in G} \alpha_x | x \rangle \mapsto 
\sum_{\rho \in \hat{G}}  \sqrt{\frac{d_{\rho}}{|G|}} \sum_{j,k=1}^{d_{\rho}} 
\left( \sum_{x \in G} \alpha_x \rho (x)_{j,k} \right) 
|\rho,j,k \rangle. \end{equation*}
For Abelian groups all dimensions $d_\rho$ are $1$, hence in that setting 
we focus only on the $\rho$. 
When generalizing this approach, where the $(j,k)$ registers are ignored, to the non-Abelian case we speak 
of \emph{weak Fourier sampling}. 
If the hidden subgroup $H$ is \emph{normal} (i.e.\ if for al $x\in G$ we have $xH=Hx$), then 
weak Fourier sampling will solve our HS Problem. 

However, in the majority of non-Abelian hidden subgroup problems, weak Fourier sampling does not provide sufficient information to recover 
the hidden subgroup. 
For example, weak Fourier sampling fails to solve the \HSP in the symmetric group 
(Grigni et al.\ \cite{gsvv04}, Hallgren et al.\ \cite{hrts03}) and the dihedral group. 
To obtain more information about the hidden subgroup, we have to focus on 
not only the $\rho$ register but also the $j$ and $k$ registers. 
Such an approach is dubbed as \emph{strong Fourier sampling}; see \cite[\S~VII.C.,\S~VII.D]{wdc-lec}. 
For some groups, it turns out that strong Fourier sampling of single registers simply fails. 
Moore, Russell and Schulman~\cite{mrs05} showed that, 
regardless of what basis is chosen, strong Fourier sampling provides insufficient 
information to solve the \HSP in the symmetric group if you restrict yourself to 
measurements on single measurements of $(\rho,j,k)$ registers. 

\subsection{Example: Dihedral/Hidden Shift Problem}
The hidden shift problem (also known as the hidden translation 
problem) is a natural variant of the hidden subgroup problem.
In the hidden shift problem, we are given two injective 
functions $f_0 : G \to S$ and $f_1 : G \to S$, with the promise that
\begin{equation*} f_0 (g) = f_1 (sg) \quad \text{for some $s \in G$}.  \end{equation*}
The goal of the problem is to find $s$, the \emph{hidden shift}. 

Consider the case of the dihedral group $D_n$, which is the group of symmetries of a $n$-sides regular polygon, 
including both rotations and reflections generated by the following two matrices
\begin{center}
\begin{tabular}{ccl}
$A= \left( 
\begin{tabular}{cc}
$\cos (2\pi/n)$ & $\sin (2\pi/n)$ \\
$\sin (2\pi/n)$ & $\cos (2\pi/n)$
\end{tabular}
\right)$ & $\quad$ & : rotation, \\[5pt] 
$B = \left( 
\begin{tabular}{cc}
$1$ & $0$ \\
$0$ & $-1$
\end{tabular}
\right)$ & $\quad$ & : reflection.
\end{tabular}
\end{center}
It is known that the dihedral group $D_n$ is equivalent to a semidirect product of 
${\Z}/n{\Z}$ and ${\Z}/2{\Z}$ denoted by 
${\Z}/n{\Z} \rtimes {\Z}/2{\Z}$. 
Roughly speaking, ${\Z}/n{\Z} \rtimes {\Z}/2{\Z}$ is the 
set of a direct product of ${\Z}/n{\Z}$ and ${\Z}/2{\Z}$ 
whose product is defined as 
\begin{equation*} (l_1, k_1) \cdot (l_2, k_2) = (l_1 + (-1)^{k_1} l_2, k_1+k_2)  \end{equation*}
for any $(l_1, k_1)$ and $(l_2, k_2)$ in ${\Z}/n{\Z} \times {\Z}/2{\Z}$. 

We define a function $F$ on ${\Z}/n{\Z} \rtimes {\Z}/2{\Z}$ 
with the property 
\begin{equation*} F(x,0) = F(x+s,1) \quad \text{for some $s \in {\Z}/N{\Z}$}. \end{equation*}
Hence, $F(\Z/n\Z,1)$ is an $s$-shifted version of $F(\Z/n\Z,0)$. 
Ettinger and H{\o}yer~\cite{eh00} showed that this hidden shift problem can be solved with only 
$(\log n)^{O(1)}$ quantum queries to the function $F$, but it remains an open problem whether this
solution can be achieved in a manner that also efficient in its time complexity.

\subsection{Pretty Good Measurement Approach to HSP}
The idea of the \emph{pretty good measurement} is borrowed from quantum optics, 
and it gives a general approach to the problem of 
distinguishing quantum states from each other. 
In a pretty good measurement, 
for a given set of possible mixed states $\{ \rho_1, \dots, \rho_M \}$, 
we use the measurement operators $\Pi_1, \dots, \Pi_M$ defined by
\begin{equation*} \Pi_i := \frac{1}{\sqrt{\sum_{j=1}^M \rho_j}} \cdot \rho_i \cdot \frac{1}{\sqrt{\sum_{j=1}^M \rho_j}}. \end{equation*}
For many hidden symmetry problems, the PGM gives 
the measurement that extracts the hidden information 
in the most efficient way possible from the states. 
It also defines a specific measurement that one can try to implement efficiently.

\section{Approaches Towards Finding New Quantum Algorithms}
Finding new quantum algorithms has proven to be a hard problem. For students and other researchers brave enough 
to nevertheless try to expand our current set of efficient algorithms, the following three approaches are suggested.  

\begin{description}
\item[Find more applications of the Abelian HSP]
The efficient quantum solution to the Abelian HSP should have more applications than we currently are aware of.  
By learning more about number theory, commutative algebra, and algebraic geometric it should be possible to discover 
computational problems in those fields that can also be solved efficiently in the Abelian HSP framework. 
\item[Find more applications for different Non-Abelian Groups]
The non-Abelian HSP for the symmetric group and the dihedral group have well-known connections to problems in graph theory and the theory of lattices. Unfortunately we do not know, at the moment, how to efficiently solve the \HSP for these groups. Find computational problems that depend on the \HSP for groups that we \emph{do} know how to solve efficiently quantum mechanically. 
\item[Find other useful Unitary transformations] 
Step away from the HSP framework and its Fourier transform all-together and look at other unitary transformations and see what computational problems are a match for other unitary transformations. 
\end{description}


\appendix
\section{Group Theory}

Groups are an important notion in algebra and they are defined as follows.  
\begin{definition}[Group]
A group is a set $G$ together with a binary operation $\circ$ on $G$ 
such that the following three axioms hold: 
\begin{enumerate}
\item[$(1)$] $(a \circ b) \circ c = a \circ (b\circ c)$ $($association$)$ 
holds for any $a, b, c$ in $G$.
\item[$(2)$] There exists an element $e$ which satisfies $a \circ  e = e \circ  a = a$ 
for any $a$ in $G$. Such element is unique and called the identity element. 
\item[$(3)$] For each $a$ in $G$, there exists an element $b$ in $G$ such that 
$a \circ b = b \circ a = e$, where $e$ is the identity element. Such an 
element is called an inverse element of $a$, denoted $a^{-1}$. 
\end{enumerate}
\end{definition}

\begin{definition}
A group $G$ is \emph{Abelian} (or \emph{commutative}), 
if its group operation commutes (i.e.\ for all $a,b\in G$ we have $a \circ b = b \circ a$). 
\end{definition}

\begin{example}
The set of all natural numbers ${\N}$ does not form 
a group with respect to the addition operation, because not all elements $x\in\N$ 
have an inverse $-x$ in $\N$. 
\end{example}

\begin{example} 
The set of all integers ${\Z}$ forms a group with respect to addition, 
since the identity element is $0$ and each element $a$ has an inverse element $-a$. 
We can easily see that the addition on ${\Z}$ is associative. 
However $\Z$ is not a group with respect to multiplication, as its 
multiplicative inverses are not elements of $\Z$. 
\end{example}

\begin{example}
Let  
\begin{align*}
M_2 ({\C}) :=& \left\{ A = 
\begin{pmatrix}
a & b \\
c & d
\end{pmatrix} 
\Big| a, b, c, d \in {\C} \right\}, \\
GL_2 ({\C}) :=& \left\{ A \in M_2 ({\C}) \ | \ \det A \neq 0 \right\}. 
\end{align*}
Note that $GL_2 ({\C})$ is a subset of $M_2 ({\C})$. 
We see that $GL_2 ({\C})$ forms a group with respect to the matrix multiplication, 
although $M_2 ({\C})$ does not form a group under that operation. 
The matrix multiplication is associative, and there exists the unit matrix $I_2$ that plays the 
role as the identity element. In $M_2 ({\C})$ however, some elements $A\in M_2(\C)$ 
do not have an inverse $A^{-1}$. 
\end{example}

\begin{thebibliography}{99}
\bibitem{ah01}
L. M. Adleman, and M.-D. Huang, \emph{Counting points on curves 
and Abelian varieties over finite fields}, 
Journal of Symbolic Computation, {Volume~32} (2001), 171--189. 
preliminary version in ANTS-II 1996. 
\bibitem{bbht}
M. Boyer, G. Brassard, P. H{\o}yer and A. Tapp, 
\emph{Tight bounds on quantum searching}, 
Fortschritte der Physik, Volume~46 (1998), 493--505. 
\bibitem{cmmp08}
D. Cheung, D. Maslov, J. Mathew and D. Pradhan, 
\emph{On the design and optimization of a quantum polynomial-time attack on 
elliptic curve cryptography}, 
Proceedings of the 3rd Workshop on Theory of Quantum Computation, 
Communication, and Cryptography, 
volume 5106 of Lecture Notes in Computer Science, (2008) pp.\ 96--104.
\bibitem{wdc-lec}
A. Childs and W. van Dam, 
\emph{Quantum algorithms for algebraic problems}, 
Reviews of Modern Physics, {Volume~82} (2010) 1--52. 
\bibitem{cl04}
R. Cleve, \emph{The query complexity of order-finding}, 
Inf. Comput., Volume~192 (2004), 162--171, 
preliminary version in CCC 2000, eprint quant-ph/9911124. 
\bibitem{cp}
R. Crandall and C. Pomerance, 
\emph{Prime Numbers: A computational perspective}, Springer-Verlag, Berlin, 2005. 
\bibitem{eh00}
M. Ettinger and P. H{\o}yer, \emph{On quantum algorithms for noncommutative 
hidden subgroups}, 
Advances in Applied Mathematics, {Volume~25} (2000), pp.\ 239--251.
\bibitem{gsvv04}
M. Grigni, L.J. Schulman, M. Vazirani and U. Vazirani, 
\emph{Quantum mechanical algorithms for the nonabelian hidden subgroup problem}, 
Combinatorica, {Volume~24} (2004), 137--154, preliminary version in STOC 2001.
\bibitem{g}
L. Grover, 
\emph{A fast quantum-mechanical algorithm for database search}, 
Proceedings of the 28th Annual ACM Symposium on Theory of Computing (STOC '96), 
1996, pp. 212--219. 
\bibitem{hrts03}
S. Hallgren, A. Russell, and A. Ta-Shma, 
\emph{The hidden subgroup problem and quantum computation using group representations}, 
SIAM Journal on Computing, {Volume~32} (2003), 916--934, 
preliminary version in STOC 2000.
\bibitem{he07}
S. Hallgren, \emph{Polynomial-time quantum algorithms for Pell's 
equation and the principal ideal problem}, 
Journal of the ACM, {Volume~54} (2007), preliminary version in STOC 2002. 
\bibitem{k05}
P. Kaye, \emph{Optimized quantum implementation of elliptic curve 
arithmetic over binary fields}, 
Quantum Information and Computation, {Volume~5} (2005), 474--491.
\bibitem{ke06}
K. S. Kedlaya, \emph{Quantum computation of zeta functions of curves}, 
Computational Complexity, {Volume~15} (2006), 1--19.
\bibitem{lw02}
A. Lauder and D. Wan, \emph{Counting points on varieties over 
finite fields of small characteristic}, 
Algorithmic Number Theory, ed. J. Buhler and P. Stevenhagen, 
Cambridge University Press, volume 44 of Mathematical Sciences Research Institute 
Publications, 2002. 
\bibitem{ln}
R. Lidl and H. Niederreiter, 
\emph{Finite Fields}, Encyclopedia of Mathematics and Its Applications, Vol. 20, 
Cambridge Univ. Press, Cambridge, 1997. 
\bibitem{mrs05}
C. Moore, A. Russell and L. J. Schulman, \emph{The symmetric 
group defies strong Fourier sampling}, Proceedings of the 46th
IEEE Symposium on Foundations of Computer Science, 2005, 
pp.\ 479--490.
\bibitem{p90}
J. Pila, \emph{Frobenius maps of abelian varieties and finding roots 
of unity in finite fields}, Mathematics of Computation, {Volume~55} (1990), 745--763. 
\bibitem{pz03}
J. Proos and C. Zalka, \emph{Shor's discrete logarithm quantum algorithm for elliptic curves}, 
Quantum Information and Computation, {Volume~3} (2003), 317--344. 
\bibitem{sc85}
R. Schoof, \emph{Elliptic curves over finite fields and the computation 
of square roots $\bmod \ p$}, Mathematics of Computation, {Volume~44} (1985), 483--494. 
\bibitem{sh}
P. Shor, 
\emph{Polynomial-time algorithms for prime factorization and discrete logarithms 
on a quantum computer}, 
SIAM Journal on Computing, Volume~26 (1997), 1484--1509.
\bibitem{wa01}
J. Watrous, 
\emph{Quantum algorithms for solvable groups}, 
Proceedings of the 33rd ACM Symposium on Theory of Computing, 
2001, pp. 60--67.
\end{thebibliography}
\end{document}